\documentclass[10pt,twocolumn,twoside,letterpaper]{IEEEtran}

\usepackage{geometry}
\makeatletter
\def\ps@IEEEtitlepagestyle{
  \def\@oddfoot{\mycopyrightnotice}
  \def\@evenfoot{}
}
\def\mycopyrightnotice{
  {\footnotesize
  \begin{minipage}{\textwidth}
  \centering
  978-1-7281-4164-0/19/\$31.00 \copyright2019 IEEE
  \end{minipage}
  }
}
\usepackage{url}
\usepackage{graphicx}
\usepackage{lineno}
\usepackage[utf8]{inputenc}

\begin{document}

\title{Testing Highly Integrated Components for the Technological 
Prototype of the CALICE SiW-ECAL}

\author{A. Irles\thanks{Manuscript received on December 20, 2019.
    
    This project has received funding from the European Union’s Horizon 2020 Research and Innovation programme under Grant Agreement no. 654168. This work was supported by the P2IO LabEx (ANR-10-LABX-0038), excellence project HIGHTEC, in the framework Investissements d\'\,Avenir (ANR-11-IDEX-0003-01) managed by the French National Research Agency (ANR). The measurements leading to these results have been performed at the Test Beam Facility at DESY Hamburg (Germany), a member of the Helmholtz Association (HGF).
    
    A. Irles is with Universit\'e Paris-Saclay, CNRS/IN2P3, IJCLab, 91405 Orsay (e-mail: irles{@}lal.in2p3.fr)} on behalf of the CALICE Collaboration.}



\maketitle

\pagenumbering{gobble}

\begin{abstract}
  A highly granular silicon-tungsten electromagnetic calorimeter (SiW-ECAL) 
  is the reference design of the ECAL for International Large Detector (ILD) concept, 
  one of the two detector concepts for the detector(s) at the future International Linear Collider. 
  Prototypes for this type of detector are developed within the CALICE Collaboration. 
  The final detector will comprise about 10\textbf{\ensuremath{^8}} calorimeter cells that have to be integrated in
  a volume of maximal 20 cm in depth. Detector components that in terms of size and channel density 
  come already close to the specifications for future large scale experiments are progressively developed. 
  This contribution will report on the performance of a new 1.2 mm thick 9-layer PCB with wirebonded ASICs 
  and comparisons with PCBs with BGA packaged ASICs will be presented. 
  A volume of about 6\textbf{\ensuremath{\times}}18\textbf{\ensuremath{\times}}0.2 cm$^3$ is available for the digital readout 
  and the power supply of the individual detector layers that feature up to 10000 calorimeter cells. 
  We will present newly developed electronic cards that meet these constraints.
\end{abstract}


\IEEEpeerreviewmaketitle


\section{Introduction}

\IEEEPARstart{T}{he next} large accelerator based particle physics experiment
will, most possibly, be an  $e^{+}e^{-}$ collider at a relatively high energy. 
Projects of different natures are currently under discussion. 
One particular example is the International Linear Collider (ILC) which has produced a technical
design report (TDR) in 2013~\cite{Behnke:2013xla}. This project offers a wide physics program
based on collisions of polarized electron and positron beams
at several centre-of-mass energies spanning between 91 GeV and 1 TeV.
Therefore, the ILC will be seen as a Higgs boson and $f\bar{f}$ factory (including the top quark pair production 
far from the production threshold). The ILC will allow, among others, to:
\begin{itemize}
\item determine with unprecedented precision
the EW couplings of the Standard Model Z-boson and Beyond Standard Model Z$'$-bosons \cite{Durieux:2019rbz,eps,Fujii:2019zll} to all fermions separating the right and left handed contribution of the couplings (thanks to the polarization);
\item inspect with unprecedented precision the Higgs-boson sector \cite{Fujii:2019zll}.
\end{itemize} 

To accomplish the ambitious physics program, two multipurpose detectors 
have been proposed: the International Large Detector (ILD) and the Silicon Detector (SiD)\cite{Behnke:2013lya}. 
Both detectors will study in detail the final states with heavy bosons (W, Z  and H), heavy quarks ($c$, $b$ and $t$) and lighter fermions.
To meet the required precision levels, these detectors will be based on the Particle Flow (PF) techniques\cite{Brient:2002gh,Morgunov:2004ed}.
These techniques require maximizing the information provided in each collision in order to 
fully reconstruct and separate all particles generated. This implies
the construction of very compact detectors with high granularity and featuring minimum dead material.
This is particularly challenging for the calorimetric systems.
The R\&D of highly granular calorimeters for future linear colliders is conducted within the CALICE collaboration \cite{calice}.
For further about PF and CALICE R\&D we refer the reader to reference \cite{Sefkow:2015hna} and references therein. 

In this document, we discuss details on the technological prototype of the silicon tungsten electromagnetic calorimeter~\cite{KAWAGOE2020162969}, SiW-ECAL.
This calorimeter is the reference design for the electromagnetic calorimeter of the ILD.
This calorimeter, together with the hadronic calorimeter of the ILD, will
be placed inside a magnetic coil that will provide 3.5-4 T. 
The baseline design consists in a detector of 20-24 radiation lengths ($X_{0}$) 
integrated in a volume of $\sim 20$ cm. In this volume, the SiW-ECAL
has to contain the active material (silicon, Si) and the absorber (tungsten, W).
The SiW-ECAL will feature $\sim 10^8$ channels, each one reading out one of the
5$\times$5 mm$^{2}$ squared cells in which the silicon sensors will be segmented in.
Due to these requirements, the SiW-ECAL design foresees that the very-front-end (VFE) electronics will be
embedded in the modules together with the silicon sensors, the PCB and the tungsten plates. Each of these detector layers will
have, in average, 10000 channels. The data acquisition will be based on self-triggering and zero suppression mode for all channels independently.
The strict space constrains leaves no space to any active cooling system between modules and therefore all heat has to be dissipated through the
structure. Therefore, the overall power consumption has to be reduced to the maximum. For that, the SiW-ECAL
will exploit the special bunch structure
foreseen for the ILC: the $e^{+}e^{-}$ bunch trains will arrive within
acquisition windows of $\sim$ 1-2 ms width separated by $\sim$ 200 ms. During the idle time, the bias currents of the electronics will be shut down.
This technique is usually called power pulsing.
In addition, only a volume of about $6\times 18 \times 0.2$ cm$^3$ is available for the digital readout 
and the power supply of the individual detector layers.
For more details on the SiW-ECAL design and past performance of its technological prototype we refer to \cite{KAWAGOE2020162969} and references therein.

In this document we focus on two new developments. 
The first one is a new 1.2 mm thick 9-layer PCB with wirebonded ASICs so-called COB (chip-on-board).
This board is introduced in Section \ref{sec:COB}.
We also present a newly developed front-end system in Section \ref{sec:DAQ}. It is based
on a new electronic card, the SL-Board, that can deal
with the large amount of channels per detector layer and that also meet the tight space constraints.
The system of control and readout of these boards is also discussed. 
The performance of these objects at beam test at DESY is discussed in Section \ref{sec:TB}. It includes 
some preliminary comparison between COB and the standard PCBs equipped with BGA packaged ASICs.

\section{Active Signal Unit (ASU)}
\label{sec:COB}

The entity
of sensors, thin PCB (printed circuit boards) and ASICs (application-specific integrated circuits) is called Active Signal Units or ASU.
Each individual ASU has a lateral dimensions of 18x18 cm$^{2}$. Four silicon wafers
are glued onto it. The ASU is equipped with 16 ASIC for the read out and features 1024 square pad sensors ({\it p} on {\it n}-bulk type) of 5.5x5.5 mm$^{2}$.
All ASUs described in this document are equipped with the version 2a of the SKIROC \cite{Callier:2011zz,Suehara:2018mqk}
(Silicon pin Kalorimeter Integrated ReadOut Chip) ASIC which has been designed for the readout of the Silicon PIN diodes.
This ASIC, as the version 2, works in self-trigger mode but still without automatic zero suppression.
Each ASU can hold up to 4 sensors of 90$\times$90 mm$^2$, each subdivided in 256 pads.
Each of the sensor pads is connected to the ASU pads through a dot of conductive glue.
The sensors are glued to the PCBs as explained in \cite{KAWAGOE2020162969}.
The bias voltage needed for the sensor depletion is provided through a conductive foil of copper and kapton
glued to the back of the sensor and connected to the high voltage through the interface card or the SL-Board (see Section \ref{sec:DAQ}). 
All previous published results of the technological prototype of the SiW-ECAL \cite{KAWAGOE2020162969,Amjad:2014tha} have been obtained
with ASUs equipped with ASICs in different plastic/ceramic packagings. 
That generation of ASUs is called as FEV. 
So far, the most compact option
were the FEV10-13 ASUs \cite{KAWAGOE2020162969,Poschl:2019ppa,yumiura}, equipped with 16 BGA packaged ASICs.
The boards have a thickness of 3.2-3.5 mm including components and connectors.
Until now, the digital readout of the FEVs was realised
with a Detector Interface Card (DIF) interfaced with an small adapter card (SMB)
for signal buffering and power regulation which is placed between
DIF and the ASUs (see \cite{KAWAGOE2020162969,Amjad:2014tha}). 
The FEV13 design included two conceptual modifications with respect the previous FEV:
the addition of two new layers (12 in total) including one extra layer to separate 
the analogue and digital power supply layers of the ASIC 
and a modification on the ASU conectivity. For the FEV13, the 4 sets of 35 pads with 1 mm pitch 
allowing for different connectivity choices ({\it i.e.} surface soldered connectors) 
were substituted by 3 flex cables connectors with 40 pins and 0.4 mm-pitch.
This required of a redesign of the SMB. Results based on the FEV13 have been presented in a poster
contribution to the NSS/MIC Symposium in 2018 \cite{yumiura}.

\subsection{Ultra thin ASU with chip-on-board (COB)}
\label{sec:COB}

\begin{figure}[!t]
  \centering
  \begin{tabular}{l}
\includegraphics[width=3.0in]{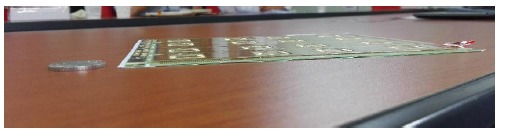}\\
\includegraphics[width=3.0in]{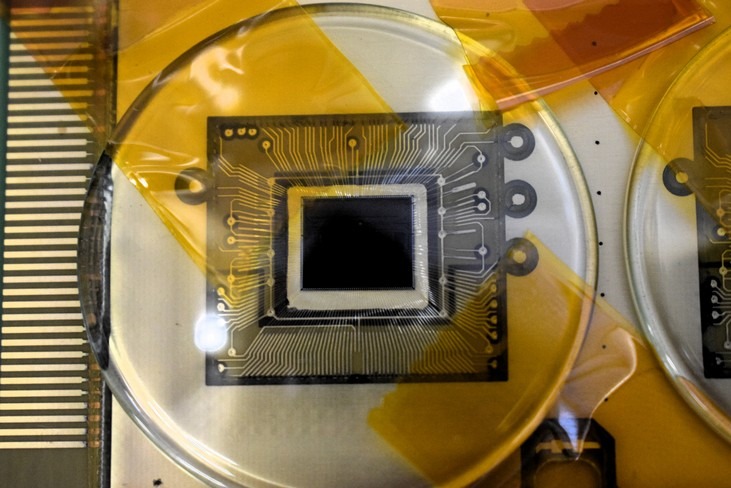}
\end{tabular}
\caption{Upper figure: one of the COB ASUs in a table, before ASIC wire bonding. This board is the result of a collaboration between the LAL and Omega institutes from France (CNRS-IN2P3) with the Sungkyunkwan University from South Korea and the EOS Corporation also from South Korea. Lower figure: detail of one Skiroc 2a wirebonded in one of the COBs. To protect the ASIC but keeping visual access to the wires, a watch glass has been added in top of it. For future productions it is foreseen that the ASICs will be protected with a synthetic resin.}
\label{cob}
\end{figure}

An alternative ASU concept has been proposed and produced. It consists on a 
ultra thin 9-layer alternative PCB design in which the ASICs
in semiconductor packaging are 
directly placed on board of the PCB in dedicated cavities and wirebonded to the PCB.
This concept of ASU is called chip-on-board ASU or COB. It has been designed to be compatible with
the batch of ASUs that were produced and operative at the time of the design (FEV10-12), {\it i.e.} same dimensions,
number and location of pads, connectors, ASICs, same high voltage supply scheme etc.
For the beam test described in Section \ref{sec:TB}, two COB boards were equipped with 
a $90\times 90$ mm Si sensor of 500 $\mu$m thickness. As for the rest of FEVs
the sensors are glued to the PCBs as explained in \cite{KAWAGOE2020162969}. 
Due to the fragility of the sensor, an excellent planarity of PCB is required in order to not damage the
sensor through mechanical stress once is glued. This is satisfied by all COBs produced
in this last batch, showing a deviation from the planarity equal to or lower than 0.5 mm.
Two photographs showing the COB are shown in 
Fig. \ref{cob}.

It is important to remark that this prototype of ASU does not foresee any space for extra components such as decoupling capacitances for the
ASIC power supplies, in contrast with the FEV generation that has up to 4 decoupling capacitances of few 100 $\mu$F per ASIC
in order to reduce disturbances in the power supplies that create spurious signals that may compromise 
the data acquisition. These signals are observed as consecutive triggers bunchs that may involve 
many channels, filling out the full memory of the ASIC quicker than real signals. They are usually denominated 
as retriggers and for more details we refer the reader to \cite{Amjad:2014tha}.

\section{Ultra compact Front End}
\label{sec:DAQ}

The foreseen space for electronic cards at the end of each detector layer to control all chained ASUs (up to 10)
and to deal with the data from the up to 10000 channels over the $\sim 2$ m of length of the layer is very limited in the ILD design.
Only 67 mm are available between the ECAL and the HCAL for those cards. In addition, a maximum height of
6-12 mm is allowed for such cards. Furthermore, such cards should have very small
power consumption levels, $<100$ mA$/$layer, so they have to be designed to work in power pulsing mode.
The solution presented here it is called the SL-Board and it has been proposed, designed and produced by the Electronics Department of the LAL.
It is described in Section \ref{sec:slboard}.
The development of this card included also the development of the front-end hardware, firmware and software
for the control and readout of the SL-Boards. The schematic view of the full 
front-end system is shown in the Fig. \ref{slboard1} and it is described in Section \ref{sec:core}.

\subsection{The SL-Board}
\label{sec:slboard}

\begin{figure}[!t]
  \centering
  \begin{tabular}{cc}
  \includegraphics[width=1.6in]{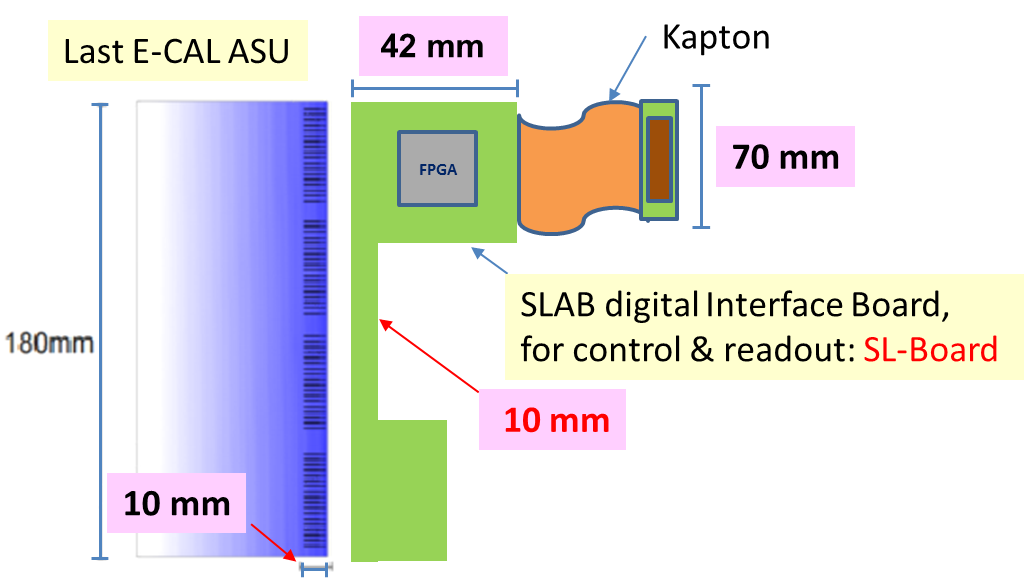}  & \includegraphics[width=1.6in]{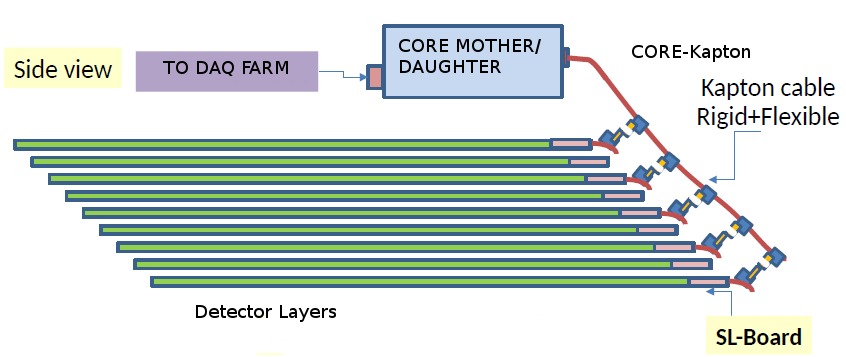} 
    \end{tabular}
\caption{Left: drawing of the spatial constraints for the control and readout electronics of the SiW-ECAL detector layers. Right: drawing of the full front-end system for one of the barrel modules of the SiW-ECAL for the ILD.}
\label{slboard1}
\end{figure}

\begin{figure}[!t]
  \centering
  \begin{tabular}{c}
\includegraphics[width=3.0in]{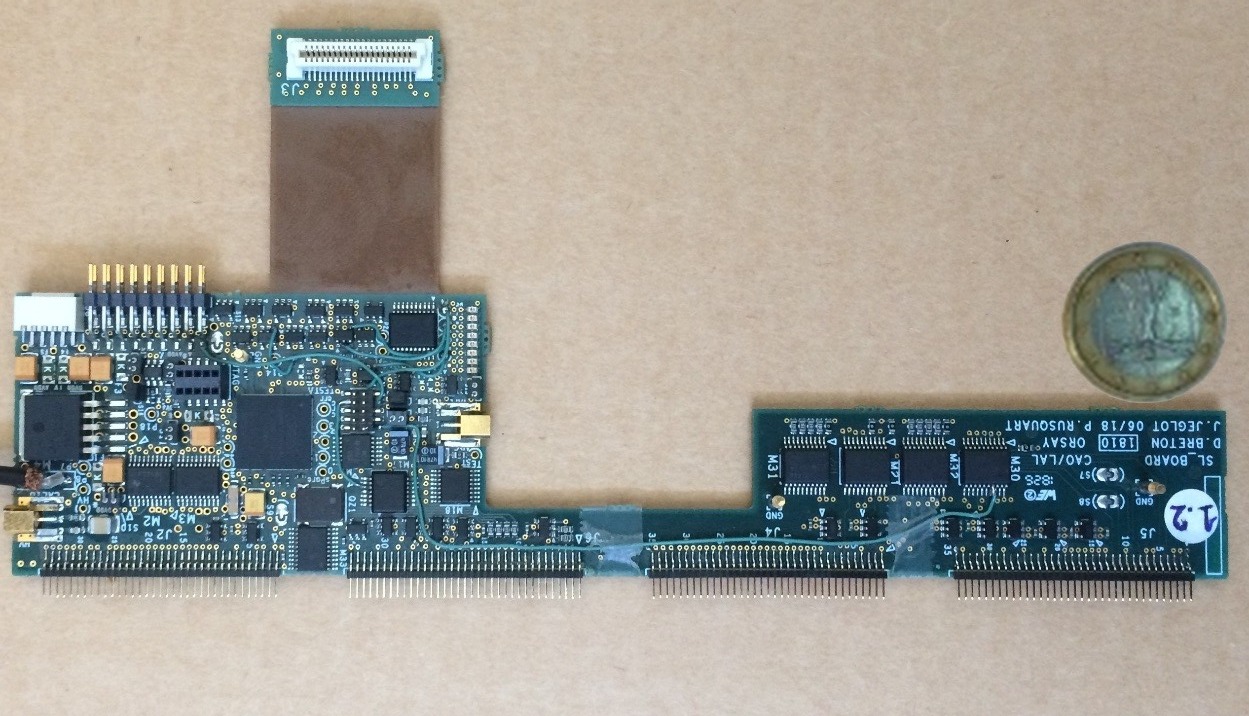}\\
\includegraphics[width=3.0in]{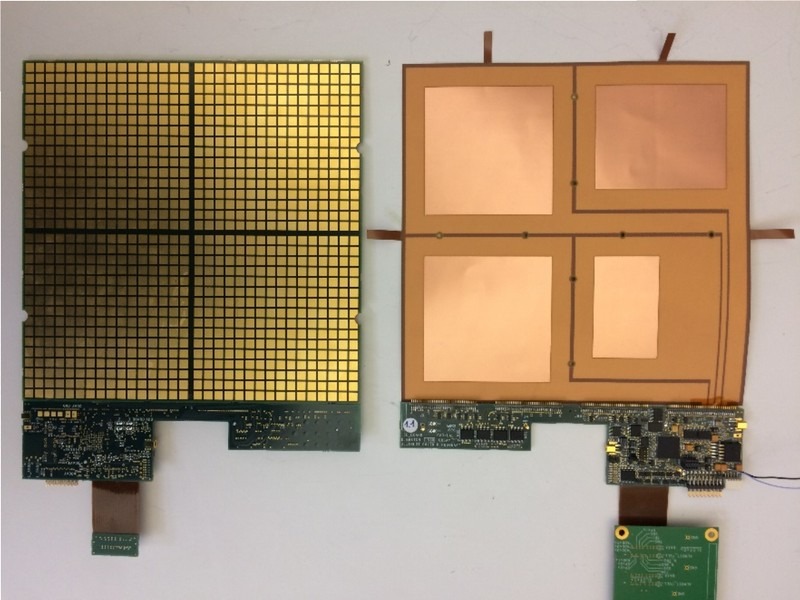}\\
\includegraphics[width=3.0in]{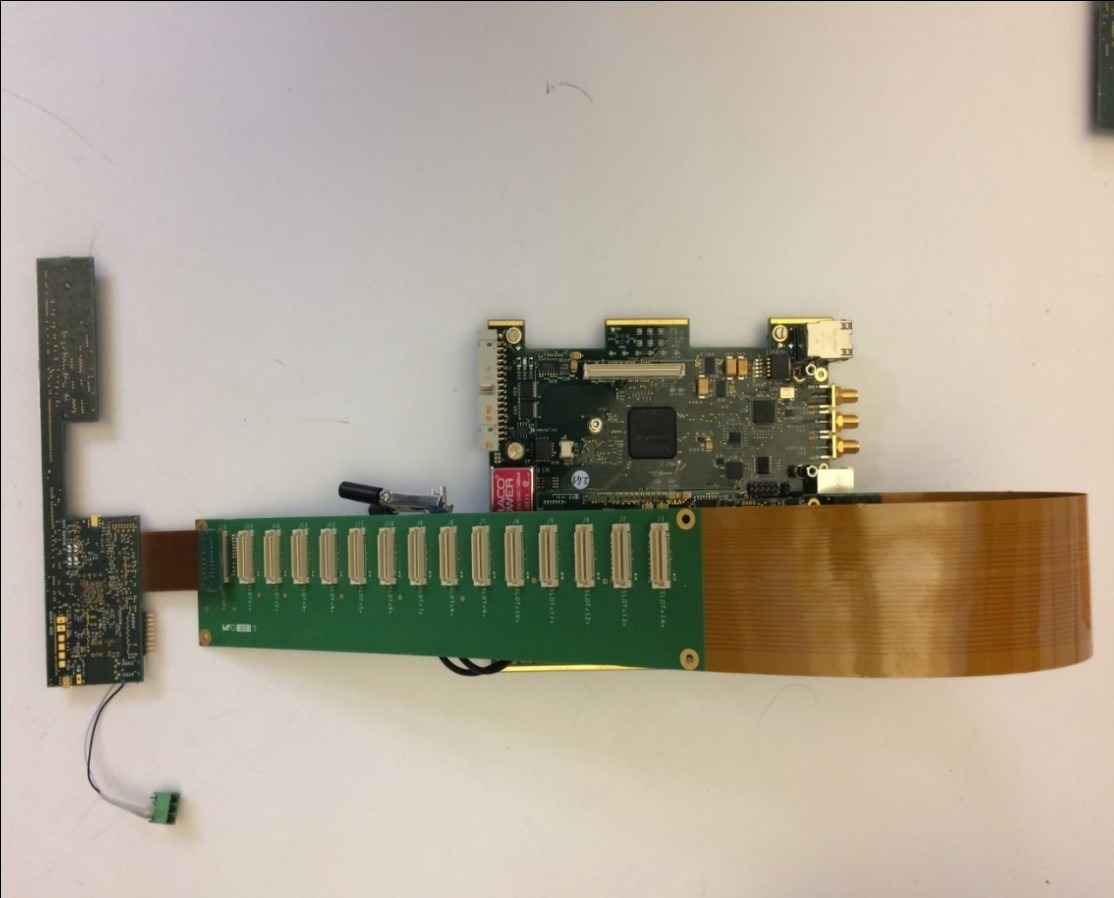} 
\end{tabular}
\caption{First figure: the first version of SL-Board tested in beam test. Second figure: an SL-BOARD connected to an ASU (backside) in the left and an  SL-Board with copper and Kapton foil used for 
the high voltage supply of the sensors. Third figure: one SL-Board connected to the CORE Mother/Daughter system through the CORE Kapton.}
\label{slboard2}
\end{figure}

The SL-Board, see Fig. \ref{slboard2}, is the sole interface for the $\sim$10000 channels of a detector layer. It is designed to
fit the space and mechanical constraints of the SiW-ECAL of the ILD (see Fig. \ref{slboard1}).
It replaces the full SMB+DIF system: the SL-Board delivers the regulated power, 
including the high voltage for the sensor bias, controls the SKIROC ASICS and performs the full data readout.
It has been designed to be fully compatible with the FEV10-12 and COB ASUs. For the FEV13, as in the case of the
SMB+DIF system, an intermediate interface is needed. This interface is under development.
The SL-Board is based on a MAX10 from ALTERA which is a mix of CPLD and FPGA and also
includes and ADC which will be used to monitor the pulsed power supply.
It is connected to the ASU through 4 surface mounted connectors of 1.5 mm height and 1 mm pitch. The SL-Board is equipped
with male connectors while the ASUs is connected with female connectors at one end and male connectors at the other end (so it can be 
plugged to another ASU in series).
The SL-Board is connected to the outside world through a system denominated CORE Kapton (see Fig. \ref{slboard2}) via an internal kapton layer and a 40-pin connector. 
This link is designed to drive, read out and synchronize up to 15 detector layers. It transmits all the clock
and fast signals and houses the control and readout links. The CORE Kapton interfaces makes use of asynchronous serial transmissions.
In addition, the SL-Board has a FTDI USB connector. It is used
for standalone and debugging tests. 

\subsection{The CORE Module}
\label{sec:core}

\begin{figure}[!t]
  \centering
  \begin{tabular}{cc}
\includegraphics[width=1.45in]{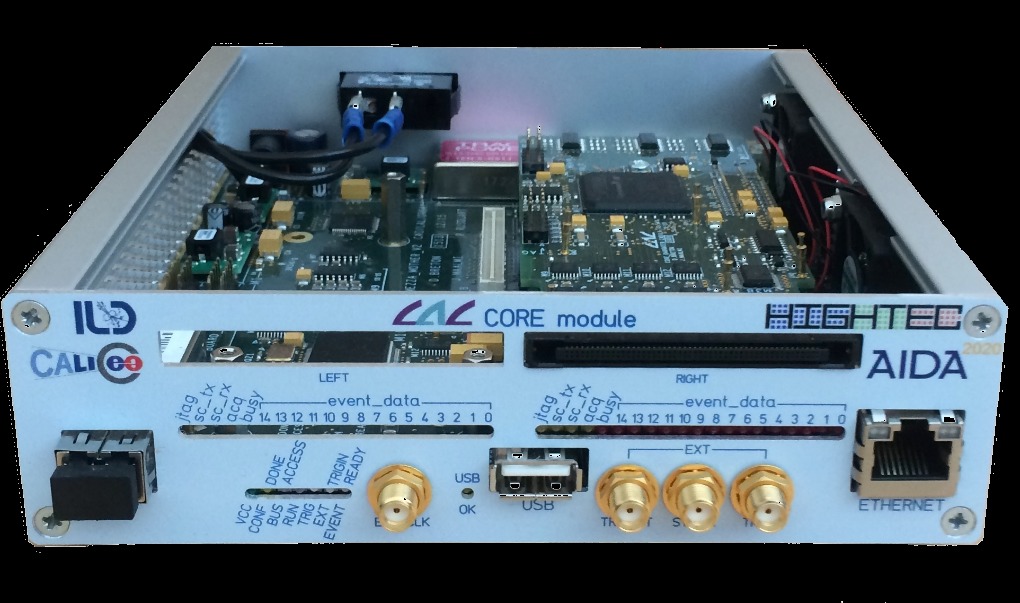} & \includegraphics[width=1.45in]{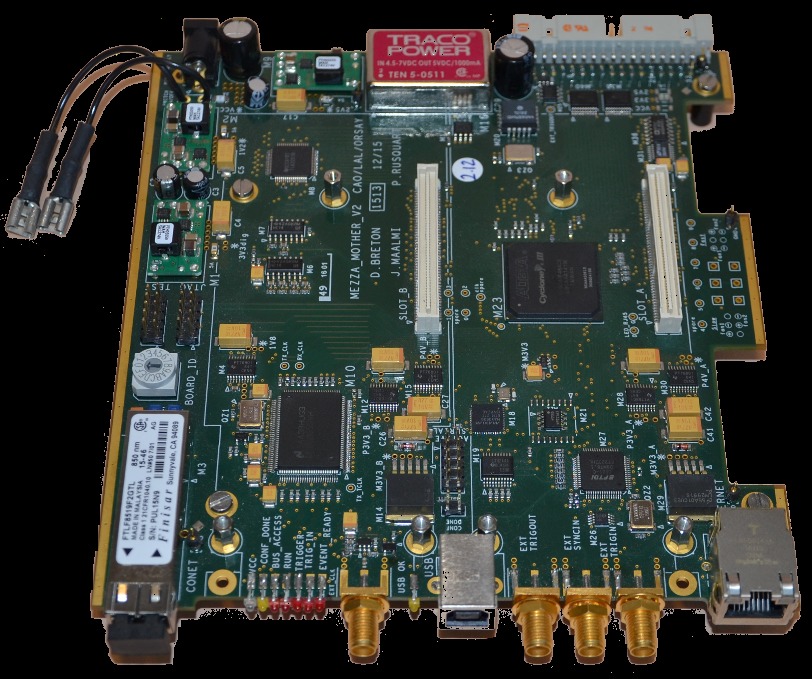} \\
\multicolumn{2}{c}{\includegraphics[width=3.0in]{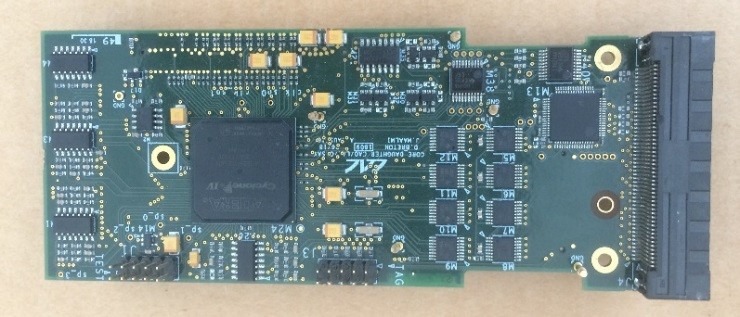}}\\
\end{tabular}
\caption{Upper figure: the CORE Mother. Lower figure: the CORE Daughter.}
\label{core}
\end{figure}

The CORE module consist of two separated pieces of hardware: the CORE Mother and the CORE Daughter. 
Both are shown in Fig. \ref{core}. The CORE Mother has been extensively used in ongoing instruments 
like the wave catcher and SAMPIC fast waveform digitizer \cite{sampic}. This control and readout motherboard has been developed for housing up to two mezzanines
and it permits separating the acquisition part from the specific front-end part. 
It manages external input and output signals for interfacing or synchronizing with other modules. 
The CORE Mother sends common clocks and fast signals
to the CORE Daughters to keep the system synchronised. The control and readout is possible
through USB (2.0), securized G-bit UDP or Optical Link (Ethernet over optical).
The CORE Daughter has been specifically developed for the SiW-ECAL prototype. It is based on a Cyclone IV FPGA. 
It is the interface between the CORE Mother and the CORE Kapton interface. It houses the second level of event buffers (derandomizers).

\subsection{Control and readout Software}
\label{sec:software}

\begin{figure}[!t]
  \centering
\includegraphics[width=3.0in]{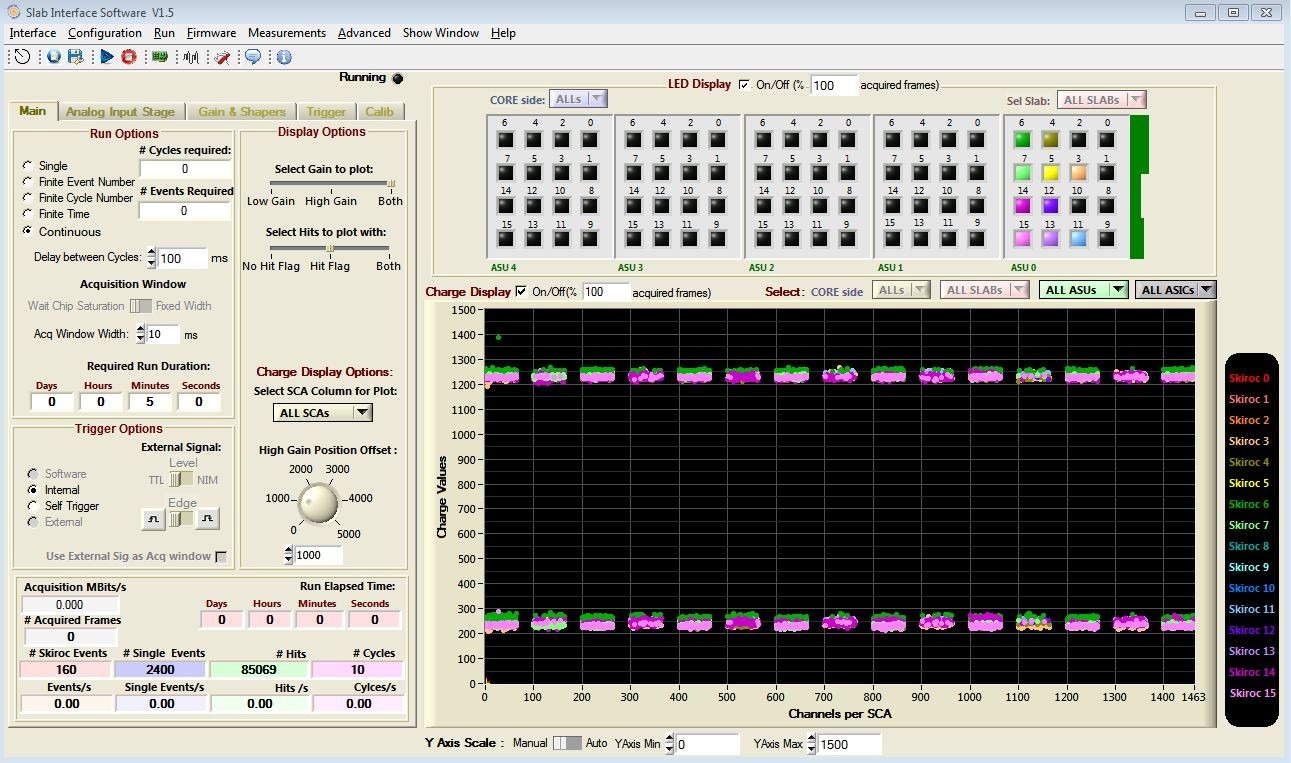} 
\caption{ Screenshot of the main window of the software interface to control and readout the modules.}
\label{soft}
\end{figure}

The acquisition software has been written in C-language and developed under LabWindows CVI. It can handle
the communication through the FTDI Module directly to the SL-Board or through the CORE Kapton. In handles
the control and readout of a whole detector module consisting on two CORE Kapton connected with 15 SL-Boards 
each and up to 5 ASUs connected in series to each SL-Board.
The C-functions that handle the communication (readout and configuration) can be used as a library with 
any other program that handles C-Language such as be EUDAQ2 \cite{Liu:2019wim} or Pyrame \cite{Magniette:2018wdz}.
The software allows also to perform online advanced commissioning measurements
as such as threshold scans or masking of noisy channels. A screenshot of the main window of the
human interface is shown in Fig. \ref{soft}.

\section{Performance in electron beam test}
\label{sec:TB}

\begin{figure}[!t]
  \centering
  \begin{tabular}{c}
	\includegraphics[width=3.0in]{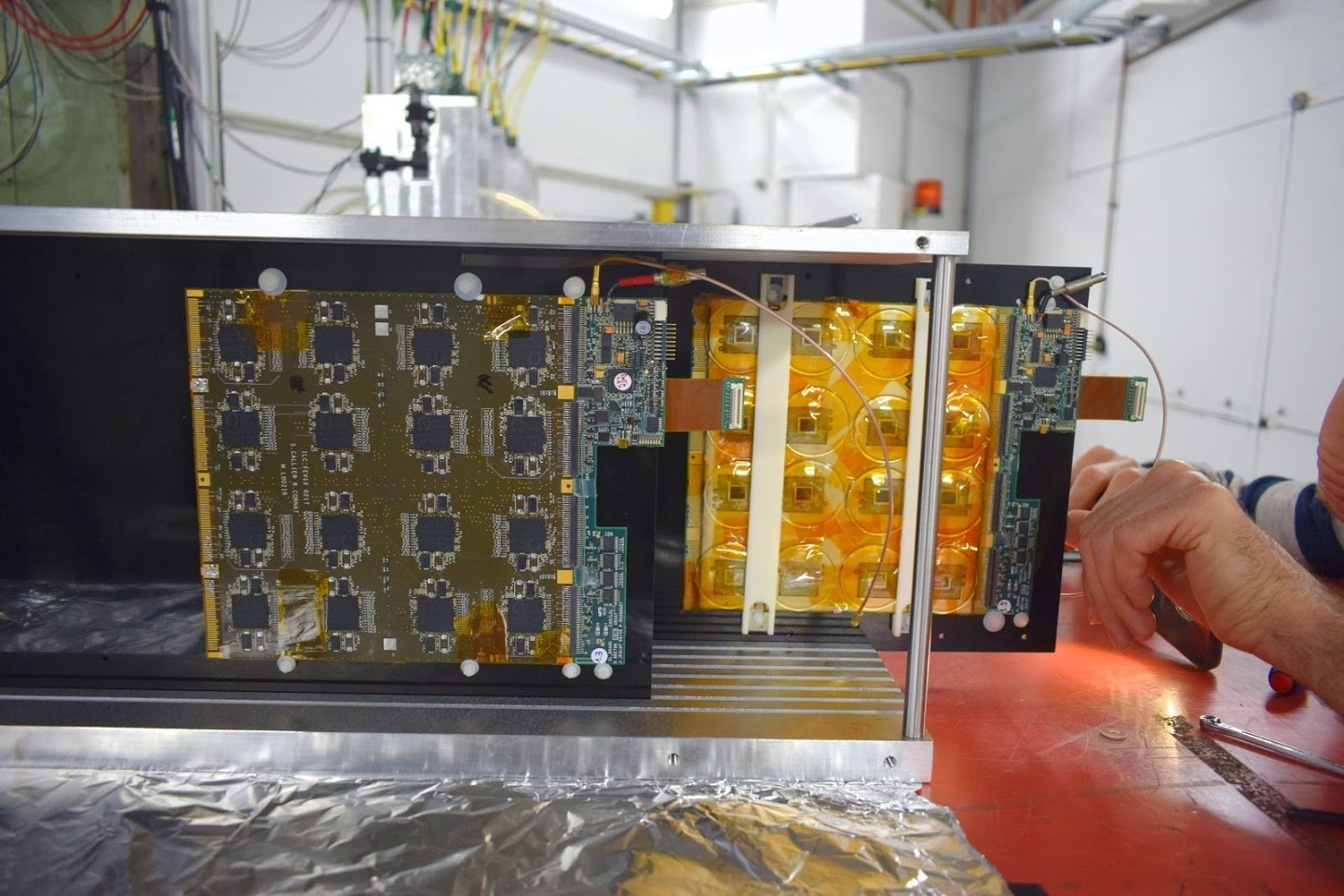} \\
	\includegraphics[width=3.0in]{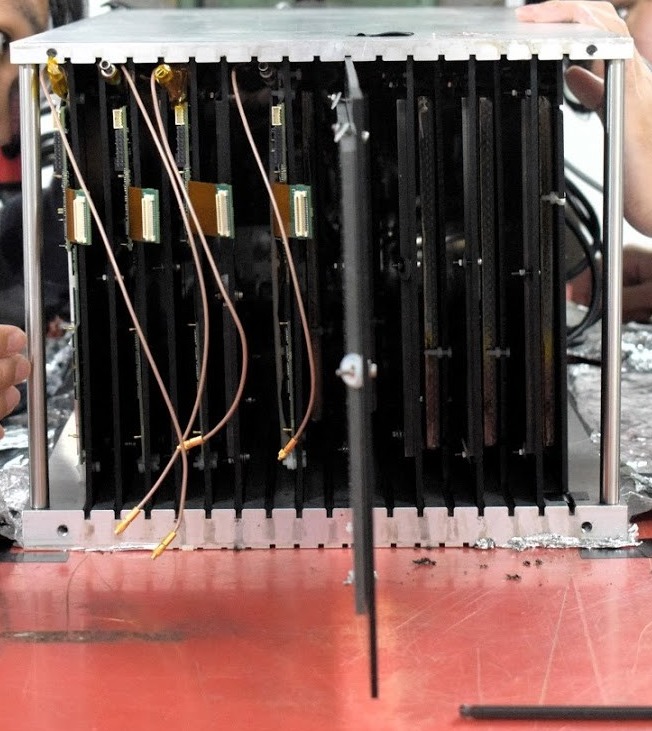} \\
	\includegraphics[width=3.0in]{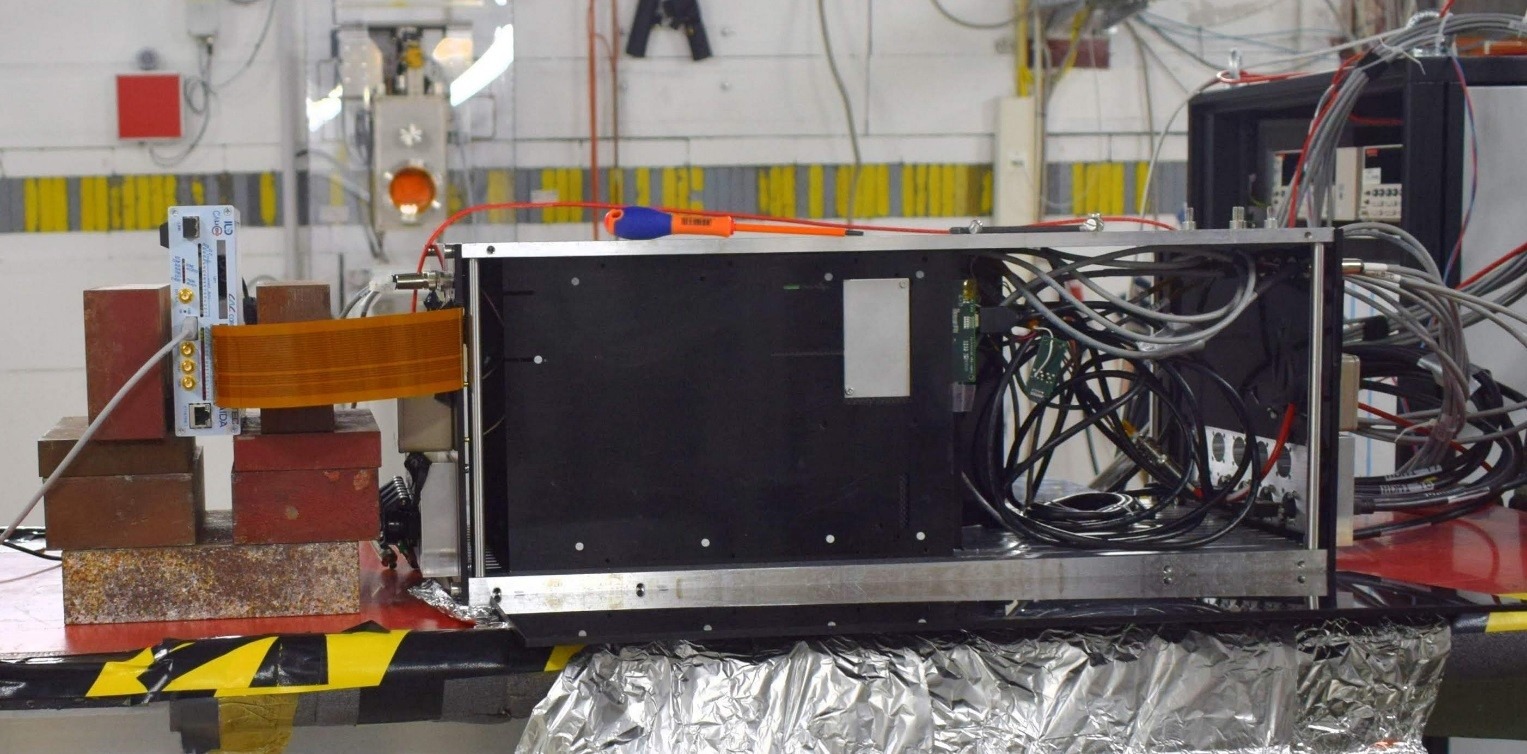} 
\end{tabular}
\caption{Top photograph: frontal view of and FEV12 and a COB ASU inside the mechanical structure. Each ASU is connected to SL-Board. Middle photograph: view of the mechanical structure from the point of view of the patch panel of the SL-Board slabs (the patch panel is removed). From right to left (beam upstream to downstream) we see 9 layers made of a plastic plate in which the ASUs plus the front end are mounted. The 5 beam upstreamer layers correspond to 5 FEV13s and the the other 4 to COB, FEV12, FEV12 and COB. Bottom photograph: frontal view of the mechanical structure. In the left we see the single CORE Kapton connector to the outside world while in the right we see the high population of cables for power and readout using the previous generation of the front end electronics. }
\label{setup}
\end{figure}

For the beam test conducted at DESY in summer 2019, up to 9 ASUs were equipped and tested. 
This was also the first beam test in which the SL-Board based front-end was tested.
They were hosted in the same mechanical structure (Fig. \ref{setup}) which was designed
to provide maximal flexibility to adapt to the different dimensions of the different modules and
also to the different front-ends. The mechanical structure, made in aluminum and plastic,
had two different patch panels: one for the SL-Board based modules and the other for the DIF+SMB based modules.
The patch panel for the SL-Board modules had 1 low voltage socket for every module, the CORE Kapton for the readout, 
a high voltage delivery box and a USB socket for every module just for debugging purposes (this was not needed during the full beam test duration).
The patch panel for the DIF+SMB modules had 2 low voltage sockets for every module, one HDMI socket (and its cable) for every module and a high voltage delivery box.
The different modules were distributed in the mechanical structure as follows (along the beam direction):

\begin{itemize}
\item Five FEV13 equipped with 4 sensors each. All these boards were readout by the previous system based on the DIF+SMB. All sensors were of 650 $\mu$m thick except the sensors in the second module that were 320  $\mu$m thick.
\item One COB connected to the CORE Kapton. This COB had, during most of the beam time, four extra decoupling capacitances (120 $\mu$F each) between the analogue power supply layer of the board and the ground. We placed these decoupling capacitances near the connector pads, since the board has no space for them near the ASICs.
\item Two FEV12 connected to the CORE Kapton.
\item Another COB connected to the CORE Kapton.
\end{itemize}

All the modules connected to the CORE Kapton were equipped with only one sensor of 500 $\mu$m each located in the closest sector to the SL-Board connectors and the MAX10.

\begin{figure}[!t]
  \centering
  \begin{tabular}{c}
	\includegraphics[width=3.0in]{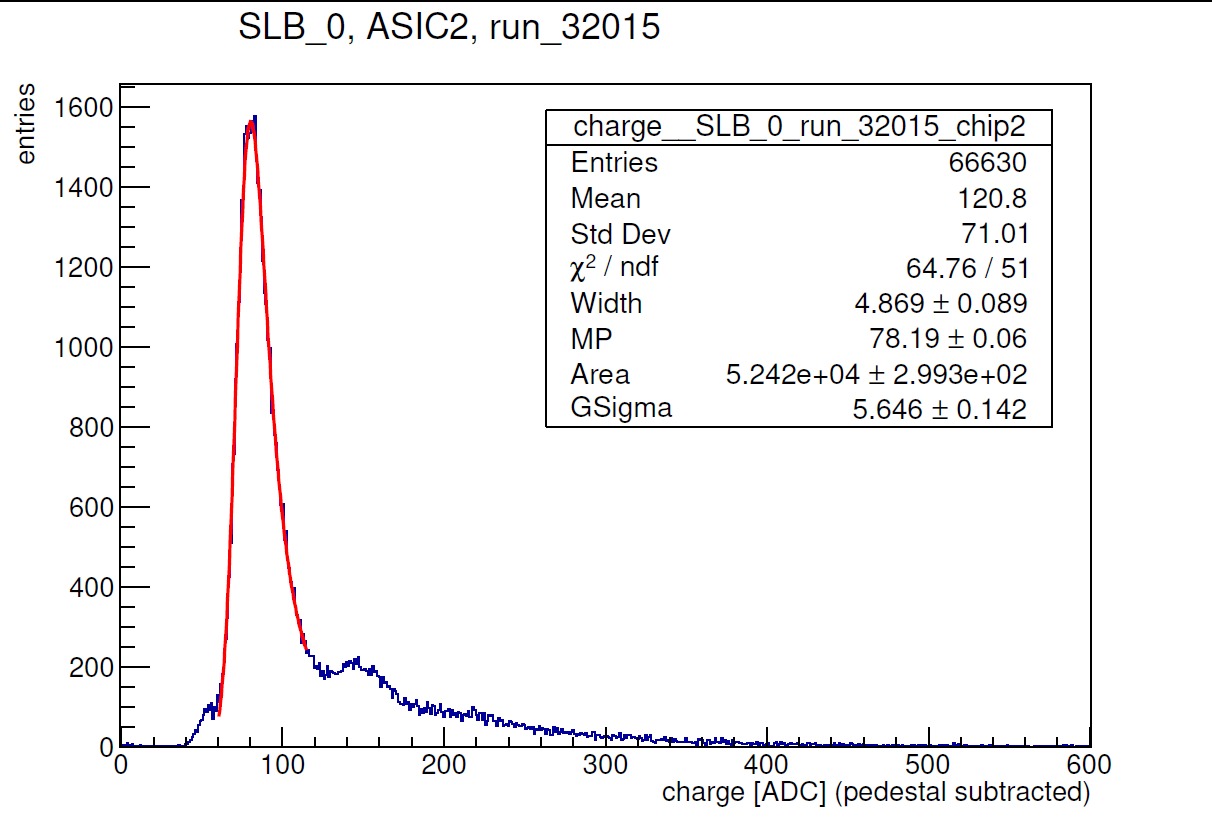} \\
	\includegraphics[width=3.0in]{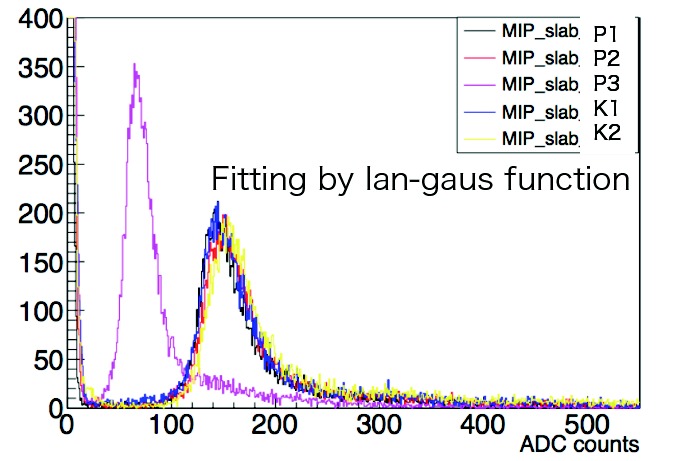}\\
	\includegraphics[width=3.0in]{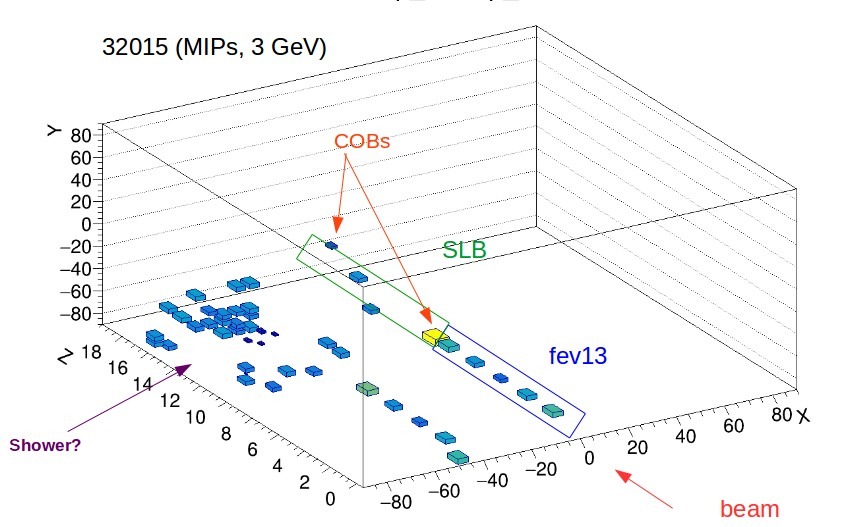}
\end{tabular}
\caption{First plot: MIP distribution for all channels readout by one ASIC of the COB boards. Second plot: equivalent distributions obtained during the same run for all the FEV13s modules. Last plot: event display, again for the same run, showing the signal of all modules for the same event. We see that two electrons
separated by $\sim$50 mm interacted with the detector at the same time. One of them even started to shower in the middle of the detector. This event display was obtained before performing any offline module alignment. }
\label{mips}
\end{figure}

The beamline at DESY provides continuous electron beams in the energy range of 1 to 6 GeV with rates from
a few hundreds of Hz to a few kHz with a maximum of $\sim$3  kHz  for  2-3  GeV.  
The full program took two weeks and most of it was dedicated to the study of the MIP response of the different
modules. This can be seen in the first two plots in Fig. \ref{mips}. There we show the MIP spectrum for
one of the COBs and the FEV13s. In both cases, the same noise filtering, pedestal correction and fit distribution procedures 
as explained in \cite{KAWAGOE2020162969} are applied. We observed that the COB in which we had added
extra decoupling capacitances had similar level of retriggers than the FEV12 and FEV13 boards, although the FEV12 and 13 are 
equipped with between 2 and 8 times more decoupling capacitances than the COB and they are located near each of the ASICs.
Detailed studies on this issue are being conducted.

During the first week, both DAQs were running independently and unsynchronized. The second week, both system
got synchronized since the CORE Mother is designed to accept external signals to define the acquisition windows. 
Therefore, a signal from the DIF+SMB system defining its acquisition window was used as input by the CORE Mother.
The two DAQ softwares were still running separately and the merging and event building of the data was done offline
since both systems run with the same clock frequency (5 MHz).
This allowed to build common events, as the one observed in the last plot of Fig. \ref{mips}.

The last part of the beam test consisted in the addition of tungsten layers between the different modules 
adding up to 7.4 radiation lengths of absorber material. We performed scan with several energies and beam positions.
The analysis of these data will be the object of a future publication.

\section{Conclusions and prospects}

In this document we have summarized the status and performance in beam
test of SiW-ECAL prototype of CALICE. 
More specifically we present for first time a new proposal for an ultra compact front-end 
DAQ based on the so-called SL-Board described in the text. 
This DAQ has been designed to meet the strict requirements of space, 
power consumption and data handling capabilities of the ILD.
It has been tested for first time in beam test with very satisfactory results.
The second new development described in this document is the ultra thin PCB 
denominated COB. This board is the result of a long R\&D process but this is the first
time to be tested in beam test. The performance of this board is very promising
and it seems to be competitive with the other type of PCBs, although more detailed
tests in a beam facility will be conducted to study it in better detail.
A new beam test is planned in March 2020 also at DESY. For this beam test, a calorimeter
of up to 15 modules will be tested using the new DAQ. For this test the SL-Board
has been upgraded modifying, among others, the localization of connectors to facilitate the access
and adding an internal ADC calibration system.
The set of 15 modules will consist of a mix of COB and BGA type ASUs, including the same FEV13s
tested in 2018 and 2019 but adapted to the new DAQ.

\appendices





\ifCLASSOPTIONcaptionsoff
  \newpage
\fi

\IEEEtriggeratref{8}
\bibliographystyle{IEEEtran}
\bibliography{IEEEabrv,../references}

\end{document}